\documentclass[namedreferences,hyperref,optionalrh]{spr-sola}
\usepackage{adjustbox}
\usepackage{caption}
\DeclareCaptionLabelSeparator{widespace}{\rule{5pt}{0pt}}
\usepackage{setspace}
\usepackage[charter]{mathdesign}
\usepackage{rotating}
\usepackage{pdflscape}
\usepackage{siunitx}
\usepackage{siunitx}
\usepackage{booktabs}
\usepackage{graphicx} 
\usepackage{color}      
\usepackage{rotating} 
\usepackage{pdflscape}
\usepackage[dvipsnames]{xcolor}
 
\hypersetup{
    final=true,
    pageanchor=true,
    colorlinks=true,
    breaklinks=true,
    linkcolor=blue,
    citecolor=blue,
    urlcolor=blue,   
    pdfpagemode=UseNone,
    pdftitle={Automatic detection of Ellerman bombs in the H-alpha line},
    pdfauthor={Faryd et al.},
    pdfsubject={Solar Physics},
    pdfkeywords={Sun: photosphere - Sun: chromosphere - Sun: activity - Methods: statistical - Methods: observational}}
\usepackage[all]{nowidow}

\chardef\us=`\_

\newcommand\arcsec{\mbox{$^{\prime\prime}$}}
\newcommand\phm{\phantom{$-$}}
\newcommand\phn{\phantom{0}}
\begin{document}

\begin{frontmatter}
\title{Automatic detection of Ellerman bombs in the H$\alpha$ line}

\author[addressref={aff1,aff2}]{\inits{A.}\fnm{Arooj}~\snm{Faryad}\orcid{0009-0008-6760-7006}}
\author[addressref=aff1,email={apietrow@aip.de},corref]{\inits{A.G.M.}\fnm{Alexander G. M.}~\snm{Pietrow}\orcid{0000-0002-0484-7634}}  
\author[addressref=aff1]{\inits{}\fnm{Meetu}~\snm{Verma}\orcid{0000-0003-1054-766X}}
\author[addressref=aff1]{\inits{T.}\fnm{Carsten}~\snm{Denker}\orcid{0000-0002-7729-6415}}

\address[id=aff1]{Leibniz-Institut für Astrophysik Potsdam (AIP), An der Sternwarte 16, 14482 Potsdam, Germany}
\address[id=aff2]{Universität Potsdam, Institut für Physik und Astronomie, Karl-Liebknecht-Straße 24/25, 14476
Potsdam, Germany}

\runningauthor{A.\ Faryad et al.}
\runningtitle{Automatic detection of Ellerman bombs in the H$\alpha$ line}

\begin{abstract}
   {Ellerman bombs (EBs) are small and short-lived magnetic reconnection events in the lower solar atmosphere, most commonly reported in the line wings of the H$\alpha$ line. These events are thought to play a role in heating the solar chromosphere and corona, but their size, short lifetime, and similarity to other brightenings make them difficult to detect.}
   {We aim to automatically detect and statistically analyze EBs at different heliocentric angles to find trends in their physical properties.}
   {We developed an automated EB detection pipeline based on a star-finding algorithm. This pipeline was used on ten high-resolution H$\alpha$ datasets from the 1-meter Swedish Solar Telescope (SST). This pipeline identifies and tracks EBs in time, while separating them from visually similar pseudo-EBs. It returns key parameters such as size, contrast, lifetime, and occurrence rates based on a dynamic threshold and the more classical static `contrast threshold' of 1.5 times the mean quiet-Sun (QS) intensity.}
  {For our dynamic threshold, we found a total of 2257 EBs from 28\,772 individual detections across our datasets. On average, the full detection set exhibits an area of 0.44~arcsec$^2$ (0.37~Mm$^2$), a peak intensity contrast of 1.4 relative to the QS, and a median lifetime of 2.3~min. The stricter threshold resulted yielded 549 EBs from 15\,997 detections, with a higher median area of 0.66~arcsec$^2$ (0.57~Mm$^2$), an intensity contrast of 1.7, and a median lifetime of 3~min. These comparisons highlight the sensitivity of EB statistics to selection thresholds and motivate further work towards consistent EB definitions.} Several long-lived EBs were observed with lifetimes exceeding one hour. While the EB intensity contrast increases towards the limb, no clear trends were found between the other EB parameters and the heliocentric angle, suggesting that the local magnetic complexity and evolutionary stage dominate EB properties.
\end{abstract}
\end{frontmatter}

\section{Introduction}
     \label{S-Introduction} 

Ellerman bombs (EBs) are small ($\approx$0.5\arcsec\ diameter) transient magnetic reconnection events \citep{Kurokawa82, Nelson2013} found at the top of the photosphere near the temperature minimum \citep{Rutten2013, Bello2013, 2014Berlicki, Liu2023}. They are closely associated with flux emergence \citep{Cheung2014} and are primarily located in the vicinity of sunspots, but they are rarely seen around plages \citep{McMatch1960}. EBs have also been found in quiet-Sun (QS) regions but at a much lower frequency \citep{Luc2016, Bhatnagar2024b}. At their formation height in the upper photosphere, EB temperature increases of 200~K to 3000~K have been reported \citep[e.g.][]{Georgoulis2002, 2014Berlicki, Hong2014, 2016Reid, 2017Fang}, and more recently up- and down-flows were detected at their locations \citep{Ichikawa2023}. Their ubiquitous presence, together with these temperature spikes, makes EBs a potential source of heating in higher layers of the solar atmosphere \citep[e.g.][]{Libbrecht2017, Chitta2024, Bhatnagar2024}, where they are associated with temperatures as high as 60\,000~K in the transition region \citep{vissers2019}. 

\begin{figure}[t]
    \centering
    \includegraphics[width=\linewidth]{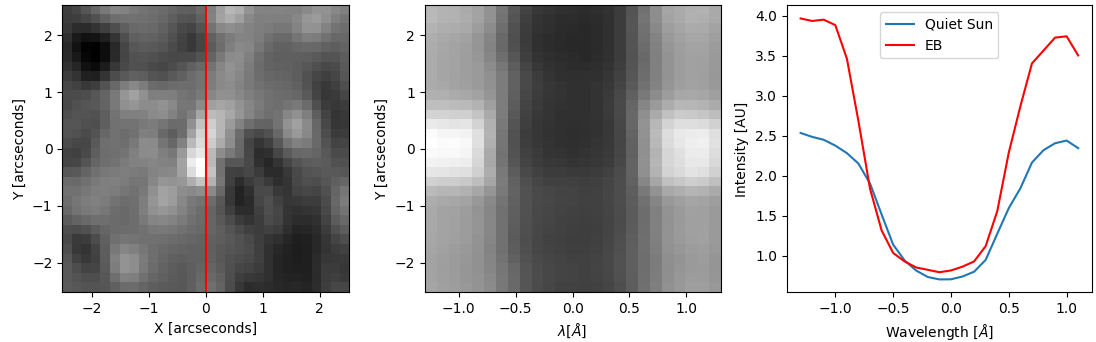}
    \caption{An example of a typical EB observation. Left: A narrowband image in the H$\alpha$ wing showing an EB with a simulated spectrograph slit (red line) passing through it. Middle: Corresponding $\lambda$--$Y$ spectrogram along the simulated slit, displaying enhanced emission in the wings (the characteristic ``mustache'' signature) caused by the EB. Right: Spectral profile extracted from the EB location (red) and the surrounding QS (blue), illustrating the enhanced wings and unchanged core. }
    \label{fig1}
\end{figure}

EBs are most commonly studied using the H$\alpha$ line, where their signature manifests itself as a strong brightening in the line wings with no effect in the line core. In fact, this is how they were first discovered by \citet{1909MitchellEB} and later expanded upon by \citet{Ellerman1917}, who called them ``solar hydrogen bombs''. Half a century later the same spectral feature was rediscovered by \citet{severny1956} who called them ``mustaches'' (see Fig.~\ref{fig1}). The spatial counterpart of the EBs was discovered by \citet{lyot1944filtre}, who called them ``petit points''. Shortly after that it was shown by \citet{McMatch1960} that the spectral brightenings (mustaches) and the spatial features (hydrogen bombs or petit points) were in fact the same physical phenomenon.

Since then, EB signatures have been found in many more spectral lines and bands, such as the other Balmer lines \citep{Joshi2022,Luc2024}, as well as the Ca\,\textsc{ii}\,H\,\&\,K lines and the Ca\,\textsc{ii} IR triplet \citep[e.g.][]{Vissers2013}, the He\,\textsc{i} 10\,830~\AA\ line \citep{Libbrecht2017}, the Fraunhofer G-band \citep[e.g.][]{Nelson2013}, and the UV \citep{Rutten2013}, including the so-called IRIS bombs \citep{Peter2014}, named after their visibility in the Interface Region Imaging Spectrograph \citep[IRIS,][]{BartIRIS2014}, and the 1600~\AA\ and 1700~\AA\ channels \citep{Qiu2000, Pariat2007, 2010Berlicki} of the Atmospheric Imaging Assembly \citep[AIA,][]{Lemen2012} on the Solar Dynamics Observatory \citep[SDO,][]{Pesnell2012}. Recently, EBs have even been detected in amateur H$\alpha$ spectral observations \citep{champeau2025}.

Despite more than a century of work on these features, many aspects of EBs remain poorly understood. This is primarily due to the limited resolving power of older ground-based telescopes and current space-based telescopes, both of which primarily see the largest and strongest EBs. This, combined with factors such as seeing, limited time series, and the confusion of other brightenings with EBs \citep[see][for a detailed discussion of this topic]{Rutten2013} has led to biases in the reported properties of EBs which have only recently begun to be addressed. For example, the average lifetime of EBs has been reported to be around 10~min to 15~min in older works \citep[e.g.][]{McMatch1960,Roy73, Kurokawa82, Qiu2000}, but more recent high-resolution studies have found it to be closer to 3~min \citep[e.g.][]{Vissers2013,2016Reid,Vissers2019A&A...627A.101V}. The same argument applies to their size, which was reported to be around 1\arcsec\ and is now believed to be 0.5\arcsec\ or below \citep{Hashimoto2010}. This creates the need for statistical studies with high cadence and high spatial and spectral resolution to re-constrain the statistics of EBs. We perform such a study based on a set of ten datasets obtained with the 1-meter Swedish Solar Telescope \citep[SST,][]{Scharmer2003} which were made publicly available by \citet{2020A&ARouppe}.

\section{Observations}

\begin{figure}[t]
    \centering
    \includegraphics[width=0.4\columnwidth,trim={1cm 1cm 1cm 0.5cm},clip]{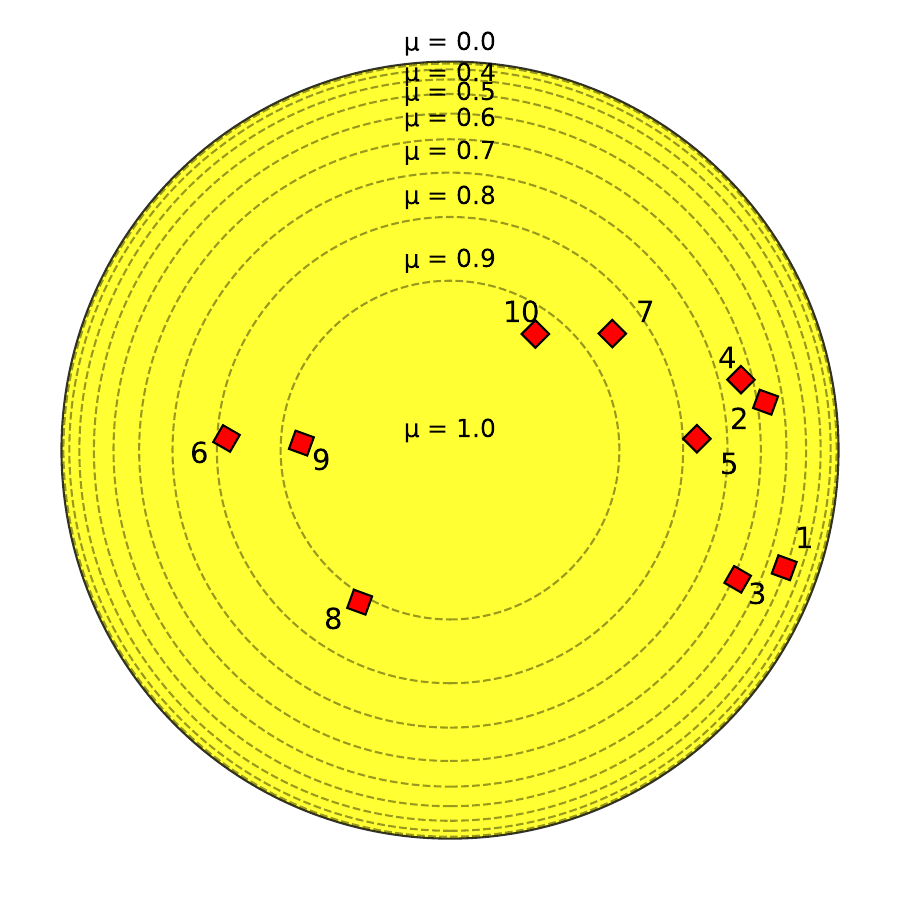}
    \caption{A schematic view of the solar disk with ten concentric rings representing the $\mu$-values between zero and unity. The location and orientation of ten SST pointings are marked and numbered.}
    \label{datset_mu}
\end{figure}

In this study, we use ten high-cadence time series, each longer than one hour. These datasets were selected from the sixteen datasets presented in~\citet{2020A&ARouppe} and are summarized in Table~\ref{tab:dataset-table}. All datasets focus on active regions at different stages of evolution and at different values of the cosine of the heliocentric angle $\mu$ as shown in Fig.~\ref{datset_mu}. 

The observations were made with the SST's CRisp Imaging SpectroPolarimeter \citep[CRISP,][]{Scharmer2008} instrument, which is a dual Fabry-Pérot interferometer that can be tuned to specific wavelengths to quickly ``scan'' through spectral lines. CRISP has an average spectral resolution of around $R \approx 130\,000$ and can observe several spectral lines in a sequence. For this work, however, we focus on the H$\alpha$ line, which has been observed with cadences ranging from 5.5~s to 20~s and a line-core sampling rate between 15 and 27 wavelength points. The pixel scale of these cubes is lower than the usual 0.058\arcsec\ pixel$^{-1}$ because they are published in a binned and cropped state to better match co-observations with IRIS. Our data have a pixel scale of 0.166\arcsec\ pixel$^{-1}$ for all datasets except dataset No.~5, which has a pixel scale of 0.133\arcsec\ pixel$^{-1}$. 

\begin{sidewaysfigure}
    \includegraphics[width=1.\textheight,keepaspectratio]{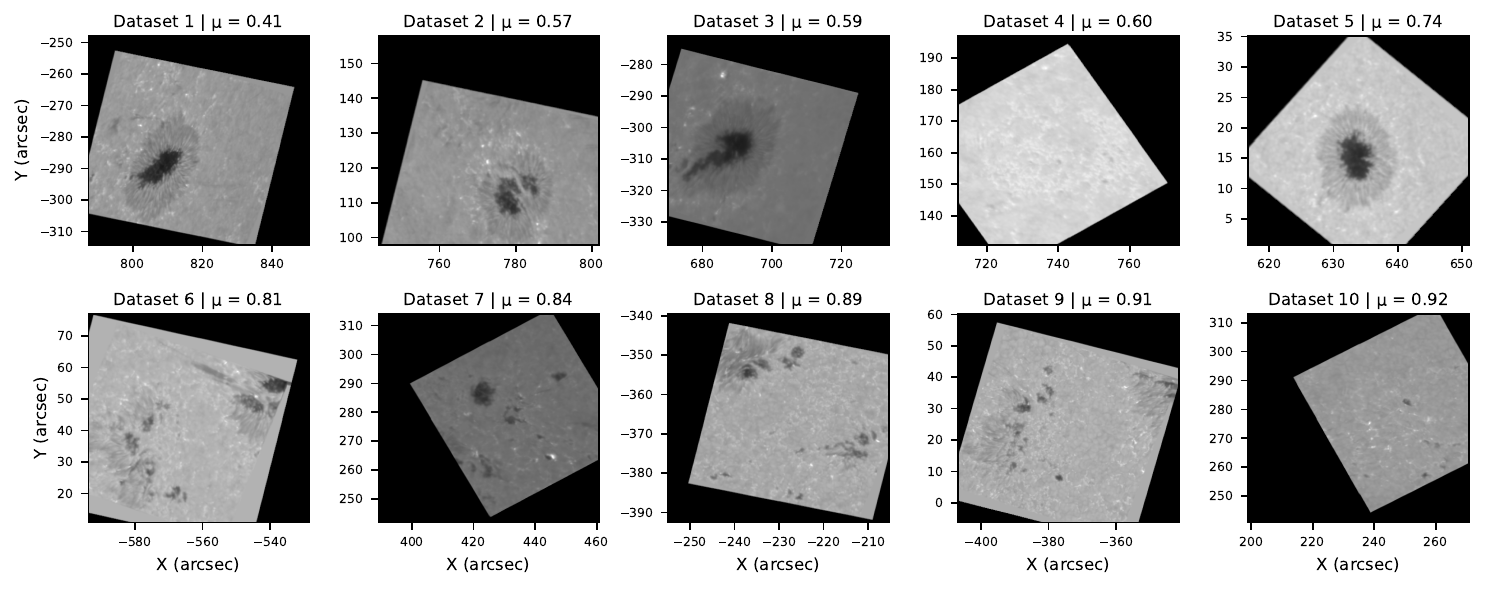}
    \caption{Images of the ten datasets used in this study, observed with the SST in different active regions and at different $\mu$-values. Each panel shows a representative image from the datasets discussed in Table~\ref{tab:dataset-table}.}
    \label{fig:datasets}
\end{sidewaysfigure}

In Table~\ref{tab:dataset-table}, we summarize key details of each set and representative images of the datasets are shown in Fig.~\ref{fig:datasets}. In the following sections, we briefly discuss each dataset.

\begin{table}[t]
\footnotesize
\caption{Observational information on the selected datasets contains the dataset number $N$, date, time, cosine of the heliocentric angle $\mu$, cadence $\delta t$, time series duration $\Delta T$, number of wavelength points $n_\lambda$, and heliocentric-Cartesian coordinates $x$ and $y$.}
\begin{tabular}{ccccccccc}
\hline
$N$ & Date & Time & $\mu$ & $\delta t$ &  $\Delta T$ & $n_\lambda$ & $x$ & $y$ \\
\hline
\phn 1 & 2014-09-06 & 08:05:37~UT & 0.41 &     11.6~s &     120.8~min & 15 & \phm 819\arcsec &      $-281$\arcsec \\
\phn 2 & 2013-09-06 & 08:11:04~UT & 0.57 & \phn 5.5~s & \phn 45.5~min & 27 & \phm 773\arcsec &    \phm 128\arcsec \\
\phn 3 & 2014-09-05 & 08:06:04~UT & 0.59 &     11.6~s &     113.1~min & 17 & \phm 702\arcsec &      $-304$\arcsec \\
\phn 4 & 2014-09-15 & 07:49:41~UT & 0.60 &     11.6~s & \phn 76.4~min & 17 & \phm 743\arcsec &    \phm 164\arcsec \\
\phn 5 & 2016-04-29 & 07:49:17~UT & 0.74 &     20.0~s & \phn 89.0~min & 17 & \phm 634\arcsec & \phm\phn 18\arcsec \\
\phn 6 & 2016-09-03 & 07:44:46~UT & 0.81 &     20.0~s &     133.3~min & 17 &   $-561$\arcsec & \phm\phn 44\arcsec \\
\phn 7 & 2014-06-15 & 07:29:31~UT & 0.84 &     11.4~s & \phn 62.5~min & 17 & \phm 425\arcsec &    \phm 278\arcsec \\
\phn 8 & 2014-09-09 & 07:59:43~UT & 0.89 &     11.6~s &     111.2~min & 17 &   $-230$\arcsec &      $-366$\arcsec \\
\phn 9 & 2016-09-04 & 07:44:46~UT & 0.91 &     20.0~s & \phn 69.0~min & 17 &   $-374$\arcsec & \phm\phn 27\arcsec \\
    10 & 2014-06-14 & 07:29:31~UT & 0.92 &     11.4~s & \phn 51.5~min & 17 & \phm 235\arcsec &    \phm 277\arcsec \\
\hline
\end{tabular}
\label{tab:dataset-table}
\end{table}

\subsection*{Dataset~1} \label{Dataset 1}

The dataset of active region NOAA~12152 was observed on September~6, 2014 at $\mu = 0.41$ with a cadence of 11.6~s and a duration of 120.8~min. This region contained a large decaying $\beta$-sunspot according to the \citet{Hale1919} classification scheme. It produced ten C-class\footnote{According to the  \citet{1970BakerGOESClassification} scale.} flares in the previous days but became magnetically simple at the time of observations. Many EBs can be visually identified around the spot. Initially, the telescope pointed at coordinates $x = 819\arcsec$ and $y = -281\arcsec$. The scan covered 15 equally spaced wavelength points in the range of $\pm$1.4~\AA\ around the H$\alpha$ line core. This dataset was previously discussed in \citet{visser2015ApJ...811L..33V}.

\subsection*{Dataset~2} \label{dataset 2}

The dataset of active region NOAA~11836 was observed on September~6, 2013 at $\mu = 0.57$ with a cadence of 5.5~s and a duration of 45.5~min. This region contained a decaying $\beta$-sunspot and produced five C-class flares during its crossing. Many EBs can be visually identified around the spot. Initially, the telescope pointed at coordinates $x = 773\arcsec$ and $y = 128\arcsec$. The scan covered 27 equally spaced wavelength points in the range of $\pm$1.2~\AA\ around the H$\alpha$ line core. This dataset was previously discussed in \citet{visser2015ApJ...812...11V}.

\subsection*{Dataset~3} \label{Dataset 3 }

The dataset of active region NOAA~12152 was observed on September~5, 2014 at $\mu = 0.59$ with a cadence of 11.6~s and a duration of 113.1~min. This region contained the same spot as seen in dataset No.~1 but one day earlier, showing rapid evolution. Many EBs can be visually identified around the spot. Initially, the telescope pointed at coordinates $x = 702\arcsec$ and $y = -304\arcsec$. The scan covered 17 equally spaced wavelength points in the range of $\pm$1.4~\AA\ around the H$\alpha$ line core. This dataset was previously discussed in \citet{visser2015ApJ...811L..33V}.

\subsection*{Dataset~4} \label{Dataset 4}

The dataset of active region NOAA~12158 was observed on September~15, 2014 at $\mu = 0.60$ with a cadence of 11.6~s and a duration of 76.4~min. This region contained primarily plage located close to a spot located outside of the FOV. No EBs can be visually identified in the region. Initially, the telescope pointed at coordinates $x = 743\arcsec$ and $y = 164\arcsec$. The scan covered 17 equally spaced wavelength points in the range of $\pm$1.4~\AA\ around the H$\alpha$ line core. This dataset was previously discussed in \citet{Skogsrud2016ApJ...817..124S}.

\subsection*{Dataset~5} \label{Dataset 5}
    
The dataset of active region NOAA~12533 was observed on April~29, 2016 at $\mu = 0.74$ with a cadence of 20.0~s and a duration of 89.0~min. This region contained a stable $\alpha$-sunspot. No EBs can be visually identified around the spot. Initially, the telescope pointed at coordinates $x = 634\arcsec$ and $y = 18\arcsec$. The scan covered 17 equally spaced wavelength points in the range of $\pm$1.4~\AA\ around the H$\alpha$ line core. This dataset was previously discussed in \citet{Bose2019A&A...627A..46B} and \citet{Drews2020A&A...638A..63D}.

\subsection*{Dataset~6} \label{Dataset 6}
    
The dataset of active region NOAA~11836 was observed on September~3, 2016 at $\mu = 0.81$ with a cadence of 20.0~s and a duration of 133.3~min. This region contained a magnetically complex region between two $\beta$-sunspots that evolved rapidly. Many EBs can be visually identified around the spot. Initially, the telescope pointed at coordinates $y = -561\arcsec$ and $y = 44\arcsec$. The scan covered 17 equally spaced wavelength points in the range of $\pm$1.5~\AA\ around the H$\alpha$ line core. This dataset was previously discussed in \citet{Nobrega-Siverio2017ApJ...850..153N} and \citet{Ortiz2020A&A...633A..58O}.

\subsection*{Dataset~7} \label{Dataset 7}

The dataset of active region NOAA~12089 was observed on June~15, 2014 at $\mu = 0.84$ with a cadence of 11.4~s and a duration of 62.5~min. This region contained several smaller pores which were part of an emerging flux region between two sunspots. Several EBs can be visually identified. Initially, the telescope pointed at coordinates $x = 425\arcsec$ and $y = 278\arcsec$. The scan covered 17 equally spaced wavelength points in the range of $\pm$1.4~\AA\ around the H$\alpha$ line core. This dataset was previously discussed in  \citet{visser2015ApJ...812...11V}.

\subsection*{Dataset~8} \label{Dataset 8} 

The dataset of active region NOAA~12157 was observed on September~9, 2014 at $\mu = 0.89$ with a cadence of 11.6~s and a duration of 111.2~min. This region contained a small flux emerging area. Several EBs can be visually identified. Initially, the telescope pointed at coordinates $x = -230\arcsec$ and $y = -366\arcsec$. The scan covered 17 equally spaced wavelength points in the range of $\pm$1.4~\AA\ around the H$\alpha$ line core. This dataset was previously discussed in \citet{Skogsrud2016ApJ...817..124S}.

\subsection*{Dataset 9} \label{Dataset 9}

The dataset of active region NOAA~12585 was observed on September~4, 2016 at $\mu = 0.91$ with a cadence of 20.0~s and a duration of 69.0~min. This region contained an emerging flux region between two rapidly evolving $\beta$-sunspots and shows cluster of pores. Many EBs can be visually identified near the penumbra. Initially, the telescope pointed at coordinates $x = -374\arcsec$ and $y = 27\arcsec$. The scan covered 17 equally spaced wavelength points in the range of $\pm$1.4~\AA\ around the H$\alpha$ line core. This dataset was previously discussed in \citet{Ortiz2020A&A...633A..58O}.

\subsection*{Dataset~10} \label{Dataset 10}

The dataset of active region NOAA~12089 was observed on June~14, 2014 at $\mu = 0.92$ with a cadence of 11.4~s and a duration of 51.5~min. This region contained a newly forming region that consists mainly of pores but will later form a flare-productive sunspot group. Some EBs can be visually identified. Initially, the telescope pointed at coordinates $x = 235\arcsec$ and $y = 277\arcsec$. The scan covered 17 equally spaced wavelength points in the range of $\pm$1.4~\AA\ around the H$\alpha$ line core. This dataset was previously discussed in \citet{visser2015ApJ...812...11V}.

\section{Methods}
    
The primary goal of our detection code is to find and track EBs throughout their lifetime. It also attempts to fill in gaps where the detection fails because of poor seeing, and attempts to identify and remove any ``pseudo-EBs'' that meet some but not all of the EB criteria laid out below. The code was designed for finding EBs in the H$\alpha$ line but can be applied to any spectral line that shows line-wing brightenings.

\begin{figure}[t]
      \centering
      \includegraphics[width=\textwidth]{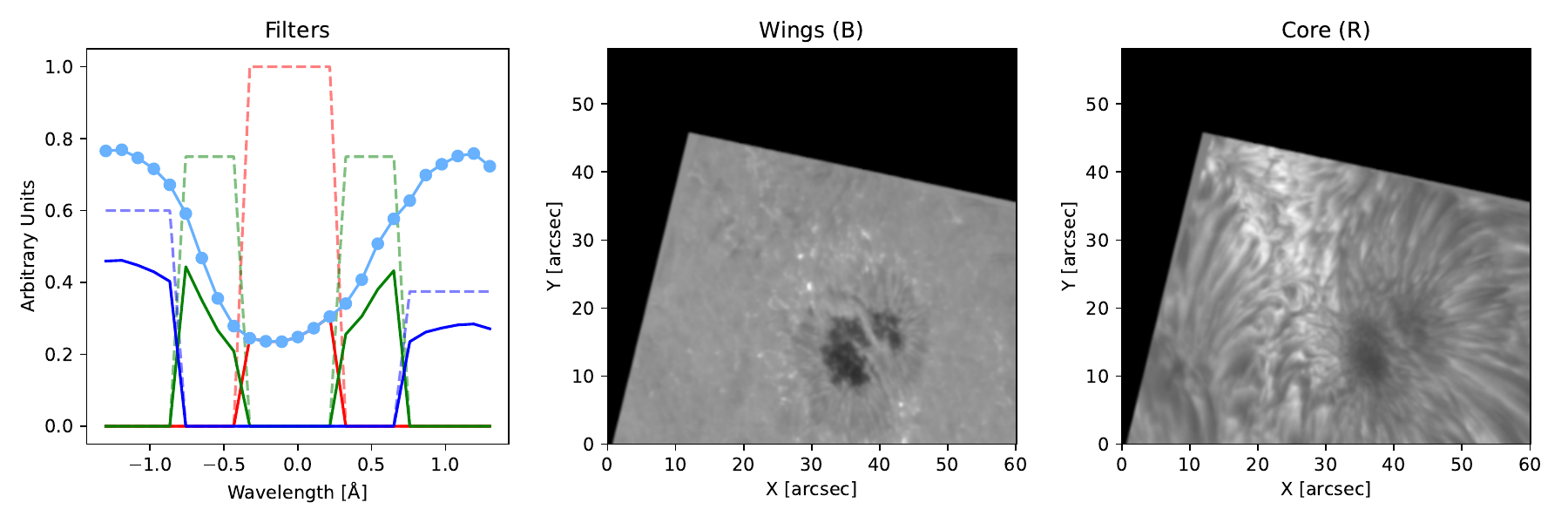}
      \caption{An example of the averaged line-wing and -core images. Left: Modified COCOPLOT filters applied to a representative H$\alpha$ profile. Three averaged images, based on the blue, green, and red filters, result from this filter. The far wings (blue) and the line core (red) are clearly separated photospheric and chromospheric components. Middle: An average line-wing image. Right: An average line-core image.}
      \label{wing&core_averahe}
\end{figure}

\subsection{Data preparation}

To detect EBs in our datasets, we used color-collapsed plotting
\citep[COCOPLOT,][]{Malcolm2021}, an algorithm that splits spectral information into three bins to produce RGB color images. We used this code to divide the spectral profile into three parts (see Fig.~\ref{wing&core_averahe}), the far line wings (blue) representing the photospheric component, the line core (red) representing the chromospheric component, and the intermediate line wings (green), which are a mixture of the two and filled with Doppler signatures. We then keep the line-core and line-wing components for further processing. In its current implementation, this is no different than averaging along certain wavelength points, but this method allows us more freedom in defining other bounds. 

Several other strategies for averaging have been tried in \citet{Faryad2025} including changing the width of each band and focusing on only one side of the profile. This is consistent with the findings of \citet{vissers2012ellerman}. We believe that this increase in EB sensitivity when using both line wings is due to the averaging of Doppler and seeing effects. Our method is similar to the one proposed by \citet{Ichikawa2023}, which is based on finding specific patterns in the contrast profiles between the QS and the investigated region on the Sun.

\subsection{Source detection} 
\label{S-simple-equations}

\begin{figure}
    \centering
    \begin{minipage}[b]{0.45\textwidth}
        \includegraphics[width=\textwidth]{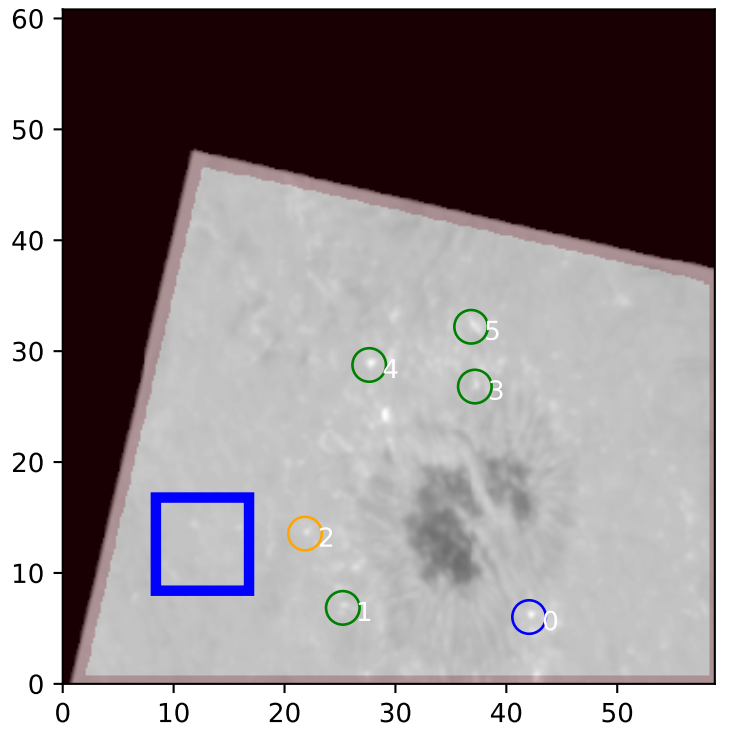}
        \centering
        {\footnotesize (a) Threshold = 7}
    \end{minipage}
    \hfill
    \begin{minipage}[b]{0.45\textwidth}
        \includegraphics[width=\textwidth]{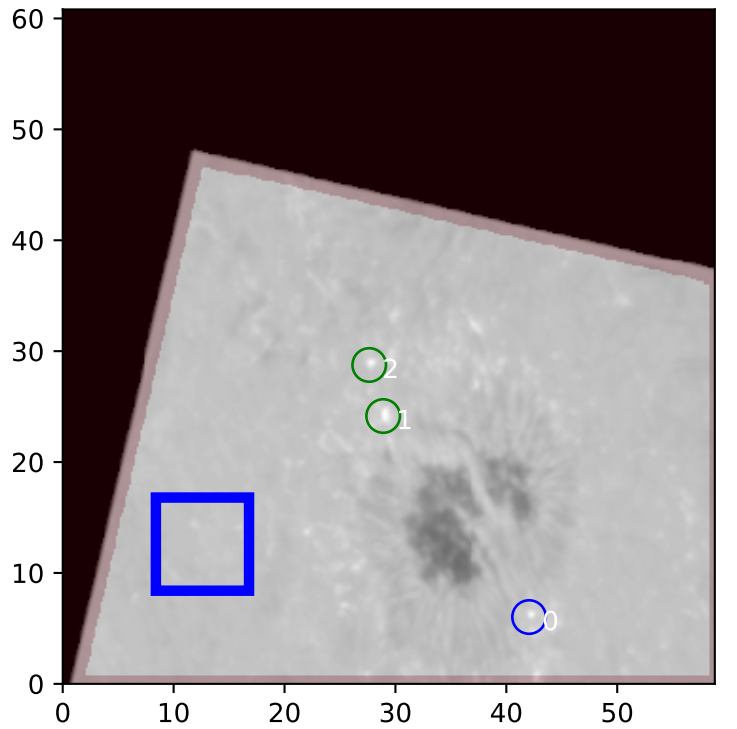}
        \centering
        {\footnotesize (b) Threshold = 10}
    \end{minipage}

    \vspace{0.5em}

    \begin{minipage}[b]{0.45\textwidth}
        \includegraphics[width=\textwidth]{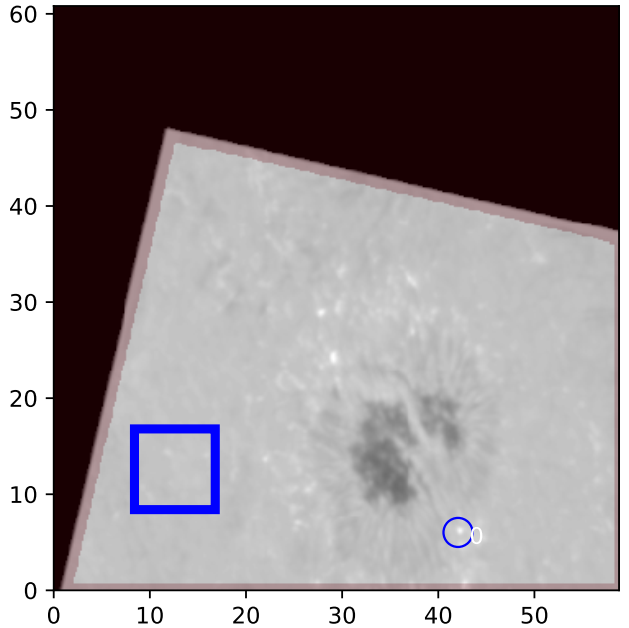}
        \centering
        {\footnotesize (c) Threshold = 13}
    \end{minipage}
    \hfill
    \begin{minipage}[b]{0.45\textwidth}
        \includegraphics[width=\textwidth]{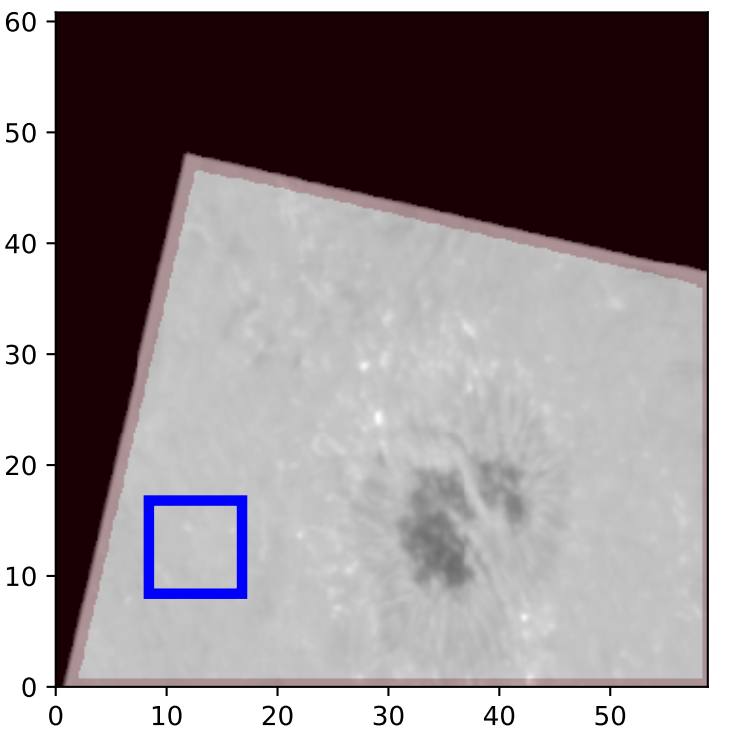}
        \centering
        {\footnotesize (d) Threshold = 15}
    \end{minipage}
    
    \caption{Example of different threshold values applied to the same frame. Lower thresholds mark more bright points, including non-EBs, while higher thresholds may miss true EB detections.}
    \label{threshold_all_images}
\end{figure}

The threshold selection process was guided by empirical testing on several frames with varying seeing conditions. After creating the line-core and line-wing averages, a QS area and threshold value are selected, with the latter being defined as the number of QS standard deviations that the detection must exceed. The QS area should be free of spots and show no significant evolution over time, while the threshold is based on the average seeing that affected the quality of the data cube. Figure~\ref{threshold_all_images} shows an example of source detections at different thresholds, with the blue square representing the QS region and the colored circles representing the source detections. The color of each circle will be explained later. 

Although no specific recipe for the optimal threshold could be found, we observed that each dataset has a relatively narrow ``sweet spot'' for this value. For thresholds below the sweet spot, the algorithm returns an order of magnitude more detections, while higher thresholds result in a sharp decrease in identified sources. This value is strongly correlated with seeing. Most datasets required a threshold around 7 $\sigma$, while those with worse seeing only worked with thresholds around~5 $\sigma$. This type of dynamic thresholding differs from most previous EB studies, which typically measure wing brightness enhancements with respect to the mean QS intensity rather than its standard deviation. To ensure compatibility with widely used criteria, we also create a subset of sources where the mean wing intensity exceeds a `contrast threshold' of 1.5 times the mean QS intensity as suggested by \citet{Rutten2013} and \citet{vissers2019}.  

\begin{figure}[t]
    \centering
    \includegraphics[width=\textwidth]{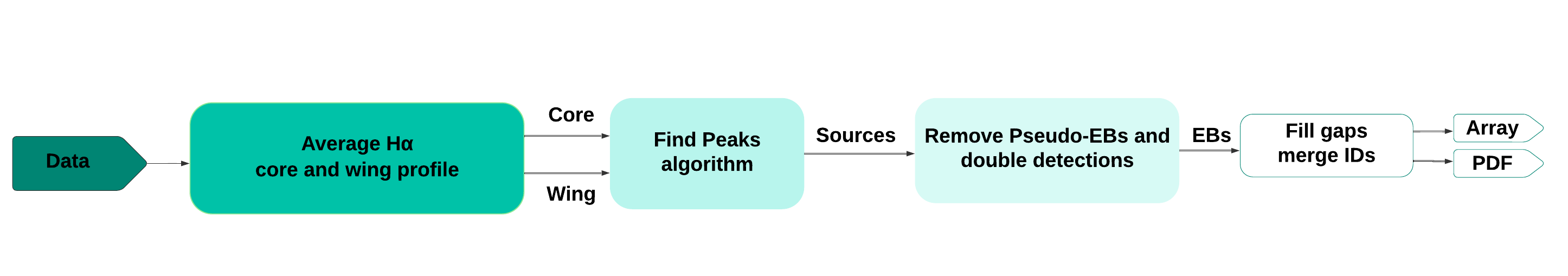}
    \caption{Flow diagram illustrating the data processing pipeline.}
    \label{Flow_code}
\end{figure}
\begin{figure}[t]
      \centering
    \includegraphics[width=\linewidth,trim={0cm 1.5cm 3cm 0},clip]{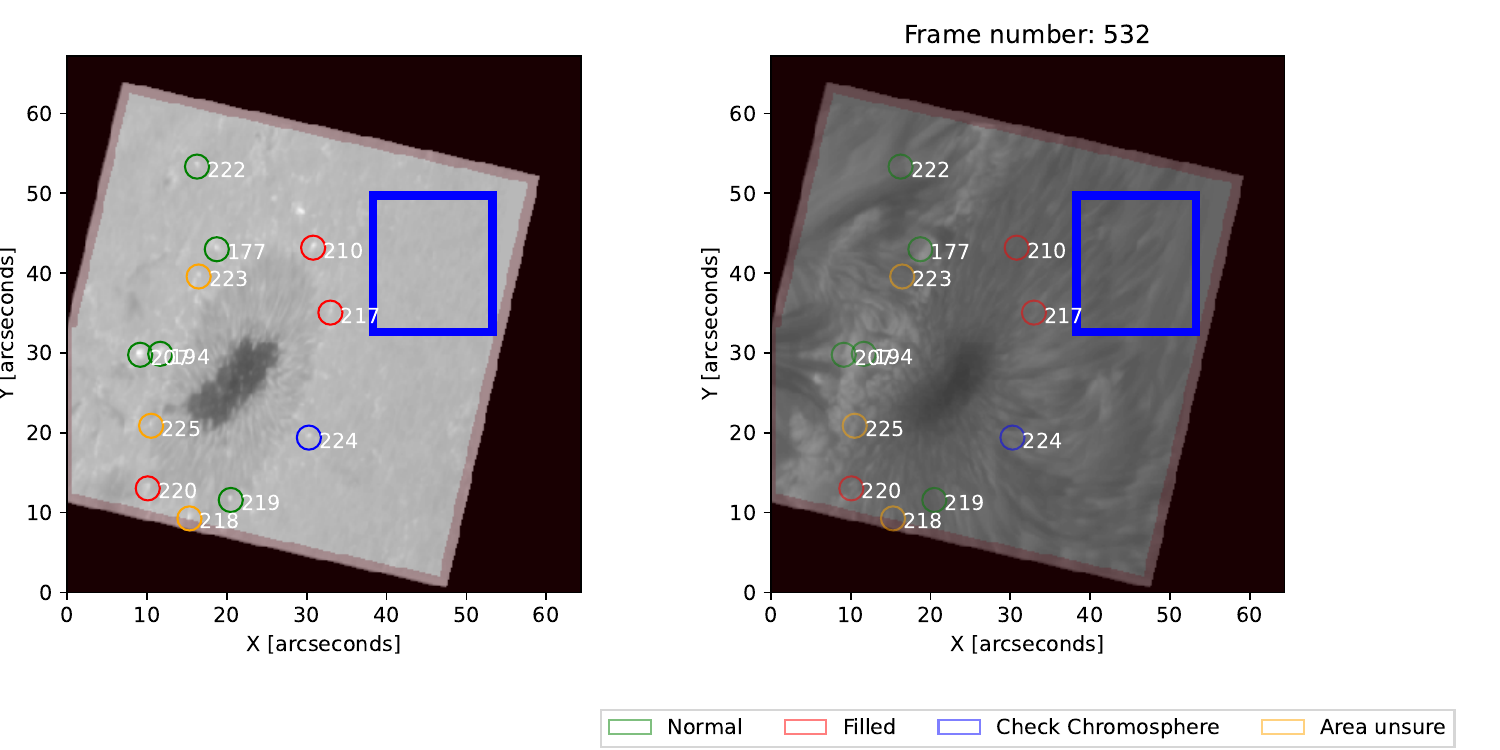}
      \caption{An example frame output of the detection code showing the FOV in the line wing and line core, with a red border that prevents edge detections, the blue square denoting the QS region, and detected EBs. Green circles denote certain detections, red detections are filled based on detections in previous and future frames, yellow detections represent uncertain area calculations, and blue detections potentially have bright counterparts in the chromosphere. }
      \label{EBs-different-visi}
\end{figure}

Our code employed two star-finding routines to search for EBs. The first is the \textit{DAOStarFinder} routine \citep{Stetson1987}. This routine finds local intensity maxima above a certain threshold that have a shape similar to a 2D Gaussian such as stars. The latter proved to be a problem because EBs are not always round, especially near the solar limb \cite[e.g.][]{watanabe2011multi}. The second routine is the \textit{Find\_Peaks} routine, which is part of the \textit{Photutils} package \citep{larry_bradley_2024_12585239}. This method works in a similar way to \textit{DAOStarFinder}, except that it has no requirements on the source shape. As a result, it can find multiple peaks in each source, but this is handled at a later point in the pipeline. The effectiveness of the two methods was compared in \citet{Faryad2025}, where the \textit{Find\_Peaks} method showed a better performance. Therefore, we will focus on this detection method. The pipeline runs the selected star-finding routine on the averaged line-wing data cube and returns an array containing the ID and coordinates of each source, which is then passed to the source processing.

\subsection{Source processing} 
\label{S-array-equations}

The resulting array is processed in four steps as illustrated in Fig.~\ref{Flow_code}. 
\begin{enumerate}
    \item For each source detection in the array, the pipeline checks for brightenings in the averaged line-core dataset. If a brightening is detected, the source is flagged as a likely pseudo-EB. EBs that are too close together (three pixels by default) are considered to be the same detection and are merged. In addition, any EBs in ``forbidden zones'' such as the image border are removed to avoid false positives caused by image edges, and detections with a diameter smaller than three pixels are rejected.
    \item Next, the pipeline loops over the array and adjusts the ID of all sources to account for those that have been removed, and to give the same ID to consecutive sources that share the same location within a given radius (three pixels by default). It also injects placeholder sources at locations where an EB disappears for a given amount of time (60~s by defeat), with their locations inferred by comparing the preceding and subsequent frames within a time window determined by the cadence of each dataset (e.g. 10 frames for a 5.5 s cadence, five frames for an 11.4/11.6 s cadence, and three frames for a 20~s cadence). It assumes that the detection was missed due to reduced contrast in poor seeing conditions. 
    \item The new resulting array is saved, and a PDF file containing context images for both the line core and line wing of each processed frame is saved (see Fig.~\ref{EBs-different-visi}). 
\end{enumerate}

\noindent The returned array contains the following quantities:
\begin{itemize}
    \item[--] \textit{ID}: unique identifier for each EB in the dataset
    \item[--] \textit{x-centroid}: $x$-coordinate of the EB in the image
    \item[--] \textit{y-centroid}: $y$-coordinate of the EB in the image
    \item[--] \textit{filled}: binary indicator indicating if the EB is injected to fill gaps in the time series
    \item[--] \textit{timeframe}: specific frame during which the observation of the EB was made
    \item[--] \textit{peak}: peak intensity of the detected EB in that frame
    \item[--] \textit{area}: estimated area occupied by the EB
    \item[--] \textit{error area}: calculated uncertainty associated with the area measurement
\end{itemize}

Several additional quantities are then calculated from this array. In this study, we focus on the quantities investigated by \citet{vissers2019}, namely the lifetime, area, contrast, and detection rate. For each EB, the lifetime is defined as the time between the first and last subsequent observation of the same EB multiplied by the observing cadence. The area of each EB is calculated after detection by fitting a 2D Gaussian to the source, resulting in an area in both arcseconds squared and megameters squared. The latter is done to minimize potential biases introduced by variations in the apparent size of the Sun due to the Earth's orbit, as well as geometrical foreshortening near the limb, while the former is retained as most EB studies report their areas in arcseconds squared. 

Here, the contrast is defined as
\begin{equation}
    C = \frac{I_\mathrm{EB}}{I_\mathrm{QS}},
\end{equation}
where $I_\mathrm{EB}$ is the peak intensity of each EB and $I_\mathrm{QS}$ is the average intensity of the QS region. The detection rate is defined as 
\begin{equation}
    D = \frac{n_\mathrm{EB}}{A_\textrm{FOV}\ \cdot \Delta T},
\end{equation}
where $n_\mathrm{EB}$ is the sum of all EB source detections, $A$ is the area of the FOV in arcseconds squared, and $\Delta T$ is the duration of the time series in minutes.


\subsection{Precision} 
\label{S-figures}

The accuracy of the detections depends on the number of false positives (FP) and false negatives (FN). We estimate this number by manual inspection, similar to \citet{Vissers2019A&A...627A.101V}. The precision is then calculated as 
 \begin{equation}
     P = \frac{\mathrm{TP}}{\mathrm{TP} + \mathrm{FP}}.
 \end{equation}
We manually inspected dataset No.~2 and found FP, FN, and $P$ to be 22\%, 15\%, and 75\%, respectively. 

\begin{sidewaystable}
\footnotesize
\caption{Observational information of the selected datasets showing results from our dynamic thresholding compared with the stricter contrast threshold. The columns show the dataset number $N$, the cosine of the heliocentric angle $\mu$, the number of individual detections $n_\mathrm{DET}$, the number of detected EBs $n_\mathrm{EB}$, the average angular area $A_\mathrm{angular}$, the metric area $A_\mathrm{metric}$, the average contrast $C$, and the detection rate $D$.}
\centering
\begin{tabular}{ccccccccc}
\toprule
{$N$} & 
{$\mu$} & 
{$n_\mathrm{DET}$} & 
{$n_\mathrm{EB}$} & 
{$A_\mathrm{FOV}$} & 
{$A_\mathrm{angular}$} & 
{$A_\mathrm{metric}$} & 
{$C$} & 
{$D$} \\
\midrule
\phn 1  & 0.41 & 5577/3927 & 332/115 & 0.92~arcmin$^2$ & 0.42/0.66~arcsec$^2$ & 0.54/0.84~Mm$^2$ & 1.52/1.67 & 2.96/1.04~arcsec$^{-2}$ min$^{-1}$ \\
\phn 2  & 0.57 & 2098/1806 & \phantom{0}55/\phantom{0}24   & 0.62~arcmin$^2$ & 0.53/0.74~arcsec$^2$ & 0.49/0.68~Mm$^2$ & 1.59/1.72 & 1.91/0.85~arcsec$^{-2}$ min$^{-1}$ \\
\phn 3  & 0.59 & 6579/6251 & 321/253 & 0.71~arcmin$^2$ & 0.48/0.53~arcsec$^2$ & 0.43/0.47~Mm$^2$ & 1.67/1.78 & 3.98/3.15~arcsec$^{-2}$ min$^{-1}$ \\
\phn 4  & 0.60 & {---}      & {---}    & 0.69~arcmin$^2$ & {---} & {---}             & {---}      & {---} \\
\phn 5  & 0.74 & {---}      & {---}    & 0.90~arcmin$^2$ & {---} & {---}             & {---}      & {---} \\
\phn 6  & 0.81 & 5624/\phantom{000} 0   & 778/\phantom{00}0   & 0.92~ arcmin$^2$ & 0.42/0.00~arcsec$^2$ & 0.27/0.00~Mm$^2$ & 1.43/0.00 & 6.31/0.00~arcsec$^{-2}$ min$^{-1}$ \\
\phn 7  & 0.84 & 1657/1625 & \phantom{0}99/\phantom{0}89   & 0.74~arcmin$^2$ & 0.97/0.99~arcsec$^2$ & 0.61/0.63~Mm$^2$ & 1.55/1.70 & 2.11/1.75~arcsec$^{-2}$ min$^{-1}$ \\
\phn 8  & 0.89 & 4125/1162 & 320/\phantom{0}25  & 0.74~arcmin$^2$ & 0.37/1.21~arcsec$^2$ & 0.22/0.72~Mm$^2$ & 1.26/1.54 & 3.87/0.42~arcsec$^{-2}$ min$^{-1}$ \\
\phn 9  & 0.91 & 1443/\phantom{000}0    & 239/\phantom{00}0   & 0.88~ arcmin$^2$ & 0.32/0.00~arcsec$^2$ & 0.19/0.00~Mm$^2$ & 1.44/0.00 & 3.89/0.00~arcsec$^{-2}$ min$^{-1}$ \\
10      & 0.92 & 1669/1226 & 113/\phantom{0}43  & 0.70~arcmin$^2$ & 0.53/0.87~arcsec$^2$ & 0.30/0.50~Mm$^2$ & 1.36/1.56 & 3.07/1.20~arcsec$^{-2}$ min$^{-1}$ \\
\bottomrule
\end{tabular}
\label{tab:results}
\end{sidewaystable}

\section{Results and discussion} 

In this study, we have analyzed ten observations of active regions at different $\mu$-values. The targets vary from those focused on sunspots, to FOVs with pores and flux emergence, to plage regions (see Fig.~\ref{fig:datasets}). Our EB detection pipeline has been applied to each of these datasets to understand the prevalence and properties of EBs under these different conditions. The code identifies and tracks EBs in each time series, allowing us to identify typical EB characteristics such as lifetime, area, and contrast.

\citet{vissers2019} used the AIA UV continuum (1600~Å and 1700~Å) to identify EBs in higher resolution H$\alpha$ images from the SST. This method gave a precision of 19\%, because many of the EBs found in the SST data were below the resolution limit of the AIA. A better precision was obtained by comparing only the 10\% brightest EBs between the two datasets. Our method detects EBs directly from the H$\alpha$ data, resulting in a much higher precision of 75\%, with a false positive rate of 22\%. While variations in seeing affect the effectiveness of the method, we find that retroactively connecting EB detections from before and after such frames helps to mitigate this. This is comparable to the recently published work by \citet{Soler2025}, who used neural networks to obtain a precision of 70\%, with only a 1\% false positive rate in SST data. However, no statistics were published to compare with.

Despite using a different detection method, we find similar results to \citet{vissers2019}. Using our variable threshold method, we find a total of 28\,772 detections (3134 of which were injected fillers), which can be grouped into a total of 2249 EBs across eight of the ten datasets presented. Using the more traditional contrast threshold, we find a total of 15\,997 detections (1791 of which were injected fillers), which can be grouped into a total of 549 EBs. This equals roughly 40\% of the detections, and 75\% of the individual EBs, as many short-lived events moved under the minimum lifetime threshold. This cut most strongly affected datasets~6 and~9, where all detections were lost. This is primarily due to the lower seeing quality of these datasets, which smear out the signals, thus lowering the peak intensity. Table~\ref{tab:results} shows the number of sources, number of EBs, average area, contrast, and detection rate per dataset. Datasets No.~4 and~5 have no detections that meet our criteria. For dataset~4, this finding is consistent with \citet{McMatch1960}, who reported that EBs are rarely seen around plages. However, the lack of EBs in dataset~5 is surprising since it is a large sunspot.  However, the seeing conditions of this spot are very bad, which smears all EB signatures to the point where they are indistinguishable from background bright points. This means that we either find no EBs or many with a very high FP rate, since every bright pixel is registered. This shows that at least variable seeing is required for this code to work.

Even excluding these two regions, we find no correlation between $\mu$ and the studied parameters, except for the contrast, which shows an expected increase with decreasing $\mu$. This implies that while the granulation contrast decreases towards the limb, the brightness of the EBs remains mostly the same. This is in line with the idea that EBs form in a narrow atmospheric region as suggested, for example, by \citet{2014Berlicki}, which would lead to a less steep limb-darkening curve \citep[e.g.,][]{Pietrow2023} However, the type of region and its evolutionary stage seems to have a much stronger influence than the $\mu$-angle, meaning that a much larger statistical study would be needed to infer any changes based on the heliocentric angle.

\begin{figure}[t]
    \centering
    \includegraphics[width=\textwidth]{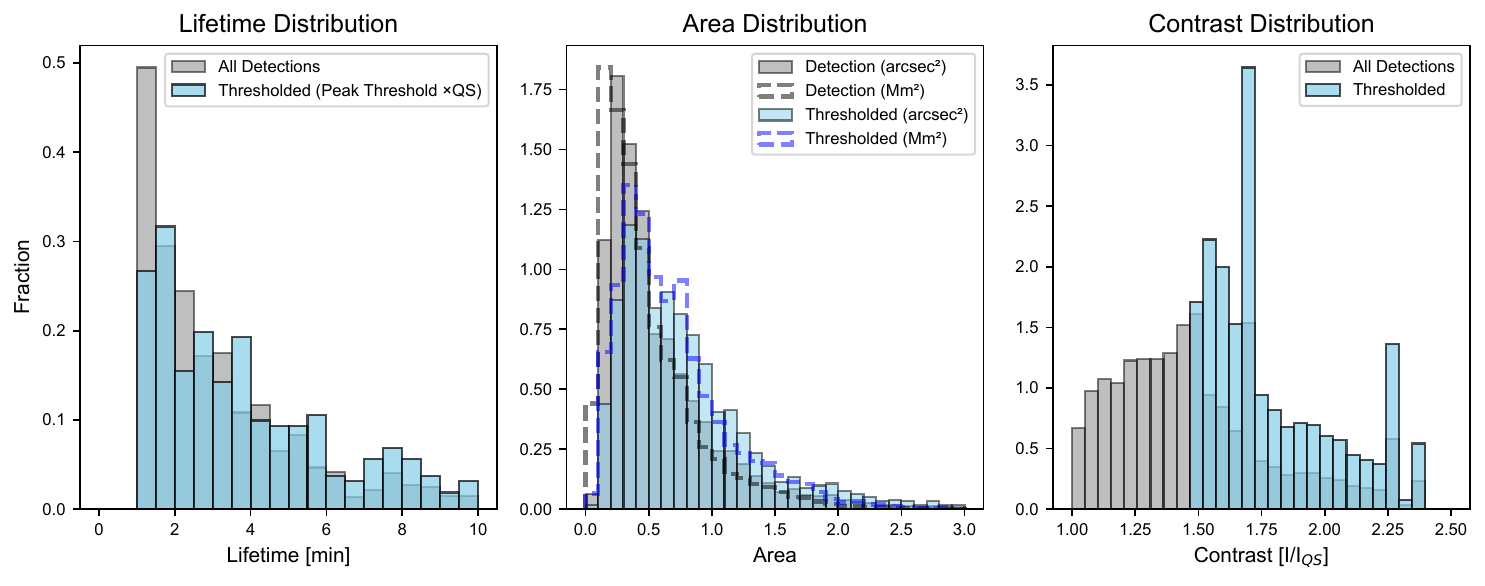}
\caption{Left: Combined histogram of EB detections across all datasets, displaying lifetime (in minutes). Results are presented for both the dynamic threshold and the contrast threshold. Middle: Histogram of EB detection area (in arcseconds squared and megameters squared) for all datasets. Right: Histogram of EB detection contrast across all datasets.}
\label{fig:results}
\end{figure}

In Fig.~\ref{fig:results}, we show histograms of the lifetime, area, and contrast for the combined EBs sample. The first panel displays the EB lifetime distribution, limited to 10~min. Within this range, we detect a total of 1200 EBs with a median lifetime of 2.33~min. When using the contrast threshold, the number of EBs detections decreases by 322 EBs within the 10~min lifetime distribution. For this sample the median lifetime increases to 3.09~min. This suggests that bright EBs tend to be longer-lived. Our two median lifetimes align closely with the average reported by \citet{Vissers2019A&A...627A.101V} and \citet{Vissers2013} but are longer than the lifetime of about 1~min suggested in Fig.~4 of \citet{2016Reid}, and shorter than the average lifetime of 7~min found by \citet{Nelson2013}. The discrepancies are likely due to different detection thresholds and sensitivities, as is shown by the roughly 1-minute difference in average lifetime between our two EB selection methods. Our more inclusive approach captures a wider range of EBs, fainter and shorter-lived brightenings that may have been missed in earlier studies.

Very long-lived (more than 1~h) EBs were reported in earlier studies \citep[e.g.][]{McMatch1960, 1972Vorpahl, Bruzek1972} and later dismissed in subsequent research as the result of low-cadence observations that failed to resolve recurrent EBs in time, leading to their misinterpretation as single, long-lived events \citep[e.g.][]{Georgoulis2002}. However, \citet{visser2015ApJ...812...11V} and \citet{Soler2025} found such events in high-resolution and high-cadence data, although it is not confirmed whether this is one long EB or a continuous series of smaller ones. We also find several such long-lived EBs which exhibit a strong oscillatory evolution, suggesting multiple reconnection events, although their flux does not fall below the detection threshold. These events require a proper investigation, which is beyond the scope of this study. 

The average area of all detections is 0.44~arcsec$^2$ (0.37~Mm$^2$) and 0.66~arcsec$^2$ (0.57~Mm$^2$) for the thresholded detections. These values are consistent with the results of \citet{Zachariadis1987} and \citet{Chen2017} but larger than the approximately 0.3~arcsec$^2$ range reported by \citet{Vissers2013} and by \citet{Nelson2013} and much larger than the 0.14~arcsec$^2$ reported by \citet{Vissers2019A&A...627A.101V} and 0.05~arcsec$^2$ by \citet{Soler2025}. This discrepancy may be partly due to the spatial resolution difference, for example, the SST data of \citet{2020A&ARouppe} used 2$\times$2-pixel binning, increasing the pixel scale from 0.058\arcsec\ pixel$^{-1}$ to 0.133\arcsec\ pixel$^{-1}$, thus leading to a coarser area detection. Since our detection method requires sources to span at least two pixels, we are by definition insensitive to brightenings smaller than 0.26\arcsec\ in diameter.

The median contrast of all detections is 1.4, increasing to 1.7 for those exceeding the peak intensity threshold. Our average contrast is thus lower than the value of 1.8 reported by \citet{Vissers2019A&A...627A.101V} but higher than the contrast of 1.29 reported by \citet{Zachariadis1985}. As with the area and lifetime measurements, this metric is sensitive to the selection criteria used for EB identification. Applying a higher intensity threshold effectively excludes weaker brightenings, resulting in a population of EBs with systematically higher contrast values.

\section{Conclusions}

In this study, we present an automated EB detection pipeline that uses high-resolution H$\alpha$ spectroscopic observations obtained with the SST. The algorithm uses averaged line-wing and line-core images to identify EBs and distinguish them from other bright features. In eight datasets, we successfully identified 2249 EBs from 28\,772 individual detections across a range of different heliocentric angles while using a dynamical threshold method. When using a more classical contrast threshold, we find 549 EBs from 15\,997 detections. With a precision of 75\%, we improve upon previous methods, but note that false positives remain a challenge in datasets with poor seeing. Furthermore, the stricter contrast threshold will reach a much lower precision rate during bad seeing conditions.

We find a median lifetime of 2.33~min for all EBs and 3.09~min for the thresholded data. Likewise, we find an average area of 0.44~arcsec$^2$ (0.32~Mm$^2$) for all EBs and 0.66~arcsec$^2$ (0.57~Mm$^2$) for the thresholded ones. Both numbers are in line with current literature, especially the ones obtained via the classical threshold. However, we remark that this contrast threshold is empirically set, and that a more dynamic seeing-correlated threshold may be more consistent. We recommend that future studies use as high spatial resolution as possible to better constrain the area of EBs values. Large telescopes with high spatial resolution, such as the Daniel K.\ Inouye Solar Telescope \citep[DKIST, ][]{2020Rimmele}, are ideal for such studies. We also recommend that future works report EB areas in physical units (megameters squared), as this may offer a more resolution-independent comparison.

As our averaged H$\alpha$ wing images are formed in a part of the line that has a relatively flat limb-darkening curve \citep{Pietrow2023}. Despite this, we find that the EB contrast has an inverse relation with the distance to the solar limb, which suggests that EBs have an even flatter limb-darkening curve. This is in line with the current understanding of EBs, which supposedly form in a relatively narrow atmospheric range.

Other parameters did not show any center-to-limb relations, suggesting that the type of active region in which the EBs form is dominant over viewing angle effects. For follow-up work, we suggest a multi-spectral approach, potentially combined with machine learning to better distinguish between EBs and EB-like phenomena. 

In addition, we report on several extremely long-lived EBs with lifetimes of more than one hour. While such events were described in older literature, they were dismissed as being the result of low-cadence and low-resolution observations, rather than physical processes. Our results challenge this assumption and should be followed up with space-based observations that are free of seeing variations.

\begin{spacing}{0.8}
\begin{acks}
The 1-meter Swedish Solar Telescope is operated on the island of La Palma by the Institute for Solar Physics of Stockholm University in the Spanish Observatorio del Roque de los Muchachos of the Instituto de Astrofísica de Canarias. This research has made use of NASA's Astrophysics Data System (ADS). DeepL Write was used in copy editing (spelling, grammar, and readability) of the manuscript.
\end{acks}
\end{spacing}
\vspace{1pt}

\begin{fundinginformation}
The Institute for Solar Physics is supported by a grant for research infrastructures of national importance from the Swedish Research Council (registration number 2017-00625). Alexander G.\ M.\ Pietrow is supported by the \emph{Deut\-sche For\-schungs\-ge\-mein\-schaft, DFG\/} project number PI 2102/1-1
\end{fundinginformation}




\begin{authorcontribution}
The individual contributions by the authors are listed according to the \textit{Contributor Roles Taxonomy} (\href{https://credit.niso.org/}{credit.niso.org}).
Arooj Faryad: formal analysis, investigation, software, validation, visualization, and writing -- original draft;
Alexander G.\ M.\ Pietrow: conceptualization, data curation, formal analysis, investigation, project administration, software, supervision, validation, visualization, and writing -- review \& editing; Meetu Verma: investigation, validation, and writing -- review \& editing; Carsten Denker: project administration, supervision, validation, and writing -- review \& editing;
\end{authorcontribution}

\begin{dataavailability}
The data are available through \citet{2020A&ARouppe} and the code is available through GitHub.
\end{dataavailability}


\bibliographystyle{spr-mp-sola}
\bibliography{ellerman}  

\IfFileExists{\jobname.bbl}{} {\typeout{}
\typeout{****************************************************}
\typeout{****************************************************}
\typeout{** Please run "bibtex \jobname" to obtain} \typeout{**
the bibliography and then re-run LaTeX} \typeout{** twice to fix
the references !}
\typeout{****************************************************}
\typeout{****************************************************}
\typeout{}}

\end{document}